\documentclass[11pt,a4paper]{article}
\usepackage{bm, amssymb,pifont,cancel,amsmath,comment,authblk}
\usepackage[dvips]{graphicx}
\usepackage{multicol}
\usepackage[colorlinks=true,link color=blue,cite color=blue,urlcolor=black]{hyperref}
\usepackage{braket}

\newcommand{\pd}{\partial}

\newcommand{\beq}{\begin{equation}}
\newcommand{\eeq}{\end{equation}}

\def\lrp#1{\left( #1 \right)}

\begin{document}
\begin{center}
\Large{\bf{Cosmological relaxation and high scale inflation}}\\ 
\vskip 6pt
\large{Tetsutaro Higaki}{\renewcommand{\thefootnote}{\fnsymbol{footnote}}\footnote[1]{E-mail address: thigaki@rk.phys.keio.ac.jp}}, \large{Naoyuki Takeda}{\renewcommand{\thefootnote}{\fnsymbol{footnote}}\footnote[2]{E-mail address: takedan@keio.jp}} and \large{Yusuke Yamada}{\renewcommand{\thefootnote}{\fnsymbol{footnote}}\footnote[3]{E-mail address: yusuke-yamada@keio.jp}}\\
\vskip 4pt
{\small{\it Department of Physics, Keio University,}}\\
{\small{\it  Kanagawa 223-8522, Japan}}\\
\vskip 1.0em
 
{\small{\it Research and Education Center for Natural Sciences,
Keio University,}}\\
{\small{\it Kanagawa 223-8521, Japan}}
\end{center}
\begin{abstract}
We study whether the relaxion mechanism solves the Higgs hierarchy problem against 
a high scale
inflation or a high reheating temperature. To accomplish the mechanism, we consider the scenario that the Higgs vacuum expectation value is determined 
after inflation. We take into account the effects of the Hubble induced 
mass and thermal one in the dynamics of the relaxion.
\end{abstract}

\section{Introduction}\label{intro}
The standard model of the particle physics~(SM) can explain 
most of results obtained from experiments.
However, the smallness of the electroweak scale is one of the theoretical mysteries when one compares it with the Planck scale.
If SM is the complete theory to describe our nature, it tells us that the nature might be unnatural because SM itself does not answer why the Higgs mass is tuned to the electroweak scale against quantum corrections, 
which are induced through the fundamental scale such as the Planck scale. 
Even if the Higgs mass is protected by some symmetry, such as supersymmetry, 
some level of tuning against the mass is 
required below the symmetry breaking scale.

Recently, a novel mechanism, called the cosmological relaxation, has been proposed in~\cite{Graham:2015cka}, which solves the hierarchy problem by the dynamics of an additional scalar field called relaxion. The relaxion has a (discrete) shift symmetry, but it is explicitly broken at a scale $M$ larger than the electroweak scale with a dimensionless coupling $g$. In this scenario, the mass of the Higgs field is 
dependent of the relaxion and is
dynamically determined by the relaxion field. Further, the relaxion potential depends on the Higgs field value 
through the instanton, i.e, through the coupling between the Higgs and 
(hidden) quark-pair condensation scale $\Lambda_c$, 
which is chiral-rotated by the relaxion-dependent phase.
By this reciprocal relation with the relaxion, the Higgs vacuum expectation value (VEV) is set to the electroweak scale in the early period of the inflation in the original literature.

The relaxion mechanism is attractive to solve the hierarchy problem, but it requires some severe constraints on the inflation scenario. To fix the field value of the relaxion by the periodic potential coming from the instanton during inflation, the Hubble expansion rate should be small. For instance, in a benchmark model, 
we find $H_{\rm inf}\sim\mathcal{O}(10^{-5})$ GeV~\cite{Graham:2015cka}, which is much smaller than the upper bound by the observations of the cosmic microwave background with $H_{\rm inf}\simeq10^{13}{\rm GeV}$~\cite{Ade:2015lrj}. 
After the inflation, inflaton decays into the SM particles and then makes thermal bath with a reheating temperature $T_{\rm R}$. Since the periodic potential is made by the non-perturbative effects through (hidden) quarks, it disappears when the chiral symmetry of the quarks are restored at a temperature higher than $\Lambda_c$. Thus, for the relaxion to be fixed even after the inflation,  naively $T_R$ needs to be lower than the condensation scale $\Lambda_c$. These requirements 
to accomplish the relaxion mechanism severely restrict the possible inflation models.

In this paper, we discuss conditions
to relax the models from such constraints exhibited
in Ref.~\cite{Graham:2015cka}, assuming that
there exists relaxion during a high scale inflation.
Therefore, we generalize the scenario by considering
various possible interaction between relaxion and other sector including 
(quantum) gravity.
Among of them, we especially take into account two effects: the Hubble induced mass and thermal one. We discuss whether the two masses change the relaxion dynamics and make the scenario compatible with the large Hubble expansion rate or high reheating temperature. In our scenario, the relaxion mechanism takes place after inflation, which is another different point from the original scenario~\cite{Graham:2015cka}.

The remaining of this paper is organized as follows. 
In Sec.~\ref{Sec2}, we review the relaxion model briefly.
In Sec.~\ref{model}, we show the relaxion model with the Hubble mass and thermal one. Then, in Sec.~\ref{dyn}, we discuss the relaxion mechanism realized after inflation and note that scenarios are classified into two ways, depending on the time of the reheating. We discuss the relaxion dynamics in each case. Finally, we summarize this paper and discuss remaining issues in Sec.~\ref{concl}.

\section{Brief review of the relaxion model}\label{Sec2}

We consider a system with a scalar field $\phi$, called relaxion. The relaxion has a discrete symmetry under $\phi\to \phi+2\pi f$. 
$f$ is the decay constant of $\phi$. 
This discrete symmetry is broken, and its explicit breaking is characterized by a dimensionless constant $g$. In this paper, we assume that a breaking into the discrete symmetry occurs during or before inflation.

The potential for the Higgs field $h$ and the relaxion $\phi$
is given with the breaking parameters of the discrete symmetry by
\begin{align}\label{V}
V=&V(g\phi/M)+(M^2-gM\phi+\cdots)|h|^2\nonumber\\
&+\lambda|h|^4+\Lambda_c^3|h|\cos\left(\frac{\phi}{f}\right),
\end{align}
where $\lambda>0$ is the Higgs quartic coupling.
$M$ is the cutoff scale of our effective field theory, and this Higgs mass should be regarded as the value renormalized above the scale $M$. 
In general, $M$ can be smaller than a fundamental scale of a theory, 
such as the Planck scale or string one. 
For example, in supersymmetric cases, 
$M$ would be the supersymmetry breaking scale.
The potential $V(g\phi/M)$ is given by series of $g\phi/M$:
\begin{align}\label{lin}
V(g\phi/M)=-M^3g\phi+\frac{1}{2}M^2g^2\phi^2+\cdots.
\end{align}
The last term in (\ref{V}) comes from the fermion condensation
caused by a hidden strongly coupled gauge theory, whose dynamical scale is 
denoted by $\Lambda_c$.\footnote{
This condensation could be identified with the SM quark condensation in QCD. 
However, in general, such a term may be originated from other hidden strong dynamics. In this paper, we only assume that the effective potential is proportional to the Higgs VEV $\langle h\rangle$.
} 
We have assumed that this term is proportional to $h$ for simplicity.
In general, however, this term can depend on $h^2$ \cite{Graham:2015cka}.
This term becomes important to fix the relaxion value in the electroweak vacuum
as discussed below. 
Note that the potential is natural in the view point of the shift symmetry~\cite{'tHooft:1979bh}: 
We then find $g \lesssim 1$ and $\Lambda_c/M \lesssim 1$.

Due to the linear potential in (\ref{lin}), the relaxion slowly rolls down toward larger field value during inflation, and after long time, $\phi$ crosses the point $\phi\sim M/g$, where the Higgs mass becomes tachyonic. Then, the Higgs field acquires the VEV, and the last term in (\ref{V}) becomes effective. By the periodic term the relaxion is fixed, equivalently the Higgs mass is stabilized at the value much smaller than the scale $M$.
This mechanism is stable against quantum corrections,
because the parameter $g$ is expected to be small through the 
naturalness argument.

Next, we focus on the condensation scale $\Lambda_c$. The hight of the periodic potential is proportional to
$4\pi f_{\pi'}^3m_N$. Here $f_{\pi'} (\sim \Lambda_c)$ is the chiral symmetry breaking scale and
$m_N $ is mass of the lightest fermion which  
is charged under the strongly coupled gauge theory. Here, $m_N < f_{\pi'}$.
For the successful relaxion mechanism, the periodic potential of the relaxion should sensitively emerge as the Higgs field value evolves to around electroweak scale.
Thus, $m_N$ is required to be sensitively determined by the Higgs VEV.
According to argument on technical naturalness in Ref.~\cite{Graham:2015cka},
we set the condensation scale to around the electroweak scale as
\beq\label{eq:scale}
\Lambda_c\sim\braket{h}\sim{\cal O} (10^2)\,{\rm GeV}.
\eeq
%

\section{Relaxion model with the Hubble and thermal masses}\label{model}

As mentioned in Introduction, we consider the relaxion model (\ref{V}) in the scenario that the relaxation of the Higgs mass takes place {\it after inflation}. This is the essential difference from the original model~\cite{Graham:2015cka}, in which the relaxation mechanism is entirely realized {\it during inflation}. 

In general, the initial value of the relaxion after inflation can be arbitrary due to the quantum fluctuation of the relaxion $\phi$, if $\phi$ is lighter than the Hubble scale during inflation $H_{\rm inf}$. Since the Higgs mass is determined by the field value of the relaxion as shown in (\ref{V}),
the fluctuations of $\phi$ leads to the inhomogeneity of the Higgs VEV among each patch of the Universe. 
In order to avoid such a problematic situation, we consider the Hubble induced mass for the relaxion given by
\begin{align}\label{Hmass}
\Delta V=\frac{1}{2}c\,H^2(\phi^2-\phi_{H}(t))^2,
\end{align}
where $c>0$ is a dimensionless constant, $H$ is the Hubble parameter, and $\phi_H$ is  a function of time $t$ in general.
If the coefficient is larger than unity as $c>1$,
the fluctuations of the relaxion during inflation are suppressed
by the Hubble induced mass. Then the relaxion value remains homogeneous
in the Universe. As a consequence isocurvature perturbation of the relaxion during inflation is suppressed \cite{Dine:2004cq}.
To realize our scenario, $\phi_H(t)$ should satisfy 
\begin{align}
\phi_{H}(t)<\frac{M}{g}
\end{align}
not to overshoot the electroweak vacuum during and after inflation. 
For simplicity, we assume $\phi_H(t)=0$ in this paper.

There are some possibilities to realize the Hubble induced mass~(\ref{Hmass})
through (quantum) gravity: A coupling between an inflaton and the relaxion 
$\frac{V_{\rm inf}}{2M_*^2}\phi^2$ gives the term~(\ref{Hmass}) with 
$c=\frac{3 M_{\rm pl}^2}{M_*^2}$ and $\phi_H(t)=0$, 
where $V_{\rm inf}$ is the inflaton potential,
and $M_{\rm pl}(\sim 2.4\times 10^{18}$GeV) is the reduced Planck scale. 
Here, $M_*$ is thought to be a scale where quantum gravity effects becomes relevant, because they would violate the global (continuous) shift symmetry \cite{Misner:1957mt}-\cite{Banks:2010zn}
in addition to the parameter of $g$.
In the string theory, $M_*$ is expected to be the string scale.
For instance, we can find $M_* \sim M_{\rm pl}/\sqrt{{\cal V}}$, 
in which ${\cal V} $ is the volume of the extra dimension in the string length unit. 
Then one finds $c \sim {\cal V} > 1$ in a large volume limit
\cite{Balasubramanian:2005zx, Higaki:2012ba}.\footnote{
If such a Hubble induced mass is generated by (stingy) instantons \cite{Blumenhagen:2009qh},
we may find the coupling $\tilde{g}V_{\rm inf}\phi^2/(2M_*^2)$, where $\tilde{g} < 1$
depends on such instanton and
may be different from $g$.
Thus, one finds $c \sim \tilde{g}{\cal V} > 1$ for large volume cases where ${\cal V} > 1/\tilde{g}$.
}
The non-minimal coupling to gravity $\xi \phi^2R$ also gives rise to the Hubble induced mass effectively: $c \sim \xi $.

Note that the Higgs quartic coupling $\lambda$ in (\ref{V}) can be negative due to the running effect, but we assume that $\lambda$ remains positive even at the inflation scale, which can be achieved with some additional scalar contribution to the running. The Higgs field $h$ also should be stabilized during inflation to avoid the inhomogeneity of the value of $h$. Such a situation can be circumvented if $M>H$ during inflation or there exists the Hubble induced mass of the Higgs field $\tilde{c}H^2|h|^2$, where $\tilde{c}>0$.

After inflation, inflaton decays into 
the SM particles. When the cosmological expansion rate drops bellow the decay rate, the reheating completes. Then, the decay products make a thermal bath with a reheating temperature $T_{\rm R}$. By the inflaton decay, the Hubble induced mass term rapidly decreases and becomes irrelevant for the dynamics of the relaxion, but alternatively the thermal bath affects the dynamics. We take into account this effect.\footnote{
Thermal effects on the relaxion model are also discussed in Ref.~\cite{Hardy:2015laa}.
}

We do not address detailed issues on the thermalization after inflation, but just add the thermal mass term of $\phi$ in an ad hoc way,
\begin{align}\label{eq:Tmass}
\Delta V=\frac{1}{2}\alpha g^2 T^2 (\phi-\phi_{T}(t))^2,
\end{align}
where $\alpha$ is a real constant\footnote{
Here, we treat $\alpha$ as a constant for simplicity. For more precise discussion of the thermal effect, we should take into account dissipation effects on the dynamics of the relaxion.
} normalized by $g^2$, 
 $T$ denotes the temperature of thermal bath, and $\phi_T(t)$ is a function of time. As in the case of the Hubble induced mass, we need to require the following condition,
\begin{align}
 \phi_T(t)<\frac{M}{g}.
\end{align}
For simplicity, we assume $\phi_T(t)=0$ in the following discussion.
As in the case of Hubble induced mass, 
the Higgs VEV is stabilized around the origin and becomes homogeneous in the presence of a temperature dependent mass term $T^2 |h|^2$ for $T \gtrsim M$.
 
In our scenario, the field value of the relaxion after inflation follows along the minimum of the potential determined by the Hubble induced mass and (or) the thermal one. When the Hubble expansion rate decreases so that $cH^2\lesssim g^2\alpha T^2$, the minimum is determined by the thermal mass. In this paper, we mainly focus on the case that the Hubble induced mass is given by a coupling to the inflaton $(V_{\rm inf}/2M_{\ast}^2)\phi^2$, then the mass term exponentially decreases as the reheating completes, $T\simeq T_R$. In this case, at $T=T_R$ the thermal mass dominates over the Hubble induced one (we assume that the temperature of the dilution plasma before reheating is sufficiently low). 
As seen below, the relaxation of the Higgs mass starts during the reheating
process where the inflaton oscillation dominates the Universe 
for a lower $T_R$, or in the radiation dominated epoch after the reheating 
for a higher $T_R$.

\section{Relaxion dynamics}\label{dyn}
We discuss the relaxion dynamics during and after inflation
in the presence of the Hubble induced mass and thermal one.
Due to the Hubble induced mass term 
$\frac{1}{2}cH^2\phi^2$, 
the relaxion $\phi$ can be stabilized at a local minimum $\phi_{\min}$,
\begin{align}
\phi_{\min}
\sim\frac{gM^2}{cH^2+g^2M^2}M
\sim \frac{gM^2}{cH^2}M . 
\label{Hdom}
\end{align}
We require the following condition during inflation,
\begin{align}
\frac{g^2M^2}{cH_{\rm inf}^2}\ll 1,
\end{align}
which ensures the validity of the expansion with respect to $g\phi$, that is, the relaxion is stabilized to a much smaller value than the critical value $\simeq M/g$.

The relaxion $\phi$ continues to stay at the minimum determined by the Hubble induced mass term even after the end of inflation, as long as 
the Hubble induced mass dominates over other contributions to the relaxion mass.
During reheating epoch, inflaton oscillates around its minimum, and the  inflaton energy density decreases as time passes. Because of this decreasing, the Hubble parameter decreases, and then $\phi$ moves toward a larger value along the time dependent minimum (\ref{Hdom}).

In our model, the relaxion mass can be dominated by the Hubble induced mass or the thermal one just before the relaxation mechanism takes place.
The potential energy of inflaton decreases soon after reheating, then
the thermal effect can dominate the relaxion mass, depending on temperature.
For a lower temperature, the Hubble induced mass may dominate it.
We classify the two cases by the reheating temperature as (A) $T_{\rm R}>\Lambda_c$ and (B) $T_{\rm R}<\Lambda_c$, and then discuss each one.

\subsection{The case (A): $T_{\rm R}>\Lambda_c$}
First, we consider the case in which $T_{\rm R}>\Lambda_c$. 
When inflation ends before the relaxion mechanism takes place,
the Hubble parameter becomes small.
Then, the Hubble induced mass term becomes ineffective for the relaxation
owing to a coupling to the inflaton.
Instead, the thermal mass term gives significant effects to the relaxion dynamics,
when the inflaton decay reheats the Universe after inflation; 
potential becomes
\begin{align}
V \sim-M^3g\phi+ \frac{1}{2}M^2g^2\phi^2+\frac{1}{2}\alpha g^2T^2\phi^2
\end{align}
for $T>\Lambda_c$. This potential is deformed adiabatically as the temperature decreases, 
and the field value in the time-dependent minimum $\phi_{\rm min}$ 
gradually increases due to the decrease of the temperature as
\begin{align}
\phi_{\rm min}\simeq \frac{M^3}{g(\alpha T^2+M^2)}.\label{Tdom}
\end{align}
The thermal mass becomes larger than the Hubble expansion rate for the relativistic degrees of freedom in the thermal bath being around $g_{\ast}={\cal O}(10^2)$ 
as in the SM and for a large $\alpha~(\gtrsim 10^2)$, which 
we will discuss below. 
Thus, the field value of the relaxion settles down to and follows along the temporal minimum (\ref{Tdom}) even in the presence of the Hubble friction.
\footnote{
The energy density of the oscillation deviated from the temporal minimum $\Delta \equiv \phi-\phi_{\rm min}$ decrease as like the radiation components, $\rho_{\Delta}\propto a^{-4}$, since the number density of the oscillator decreases by $a^{-3}$ with the decreasing of the thermal mass by $a^{-1}$. On the other hand, the kinetic energy of $\phi_{\rm min}$ does not decrease during radiation dominated epoch as $(\dot{\phi}_{\rm min})^2\sim M^6 H^2/(g^2\alpha^2 T^4)
\sim{\rm const}$. Thus, the energy of the oscillation becomes sub-dominant after sufficient expansion of the Universe.
}

As $T$ becomes lower, $\phi$ goes down to the larger field value, and finally reaches the critical value 
\begin{align}
\phi_c=\frac{M}{g}
\end{align}
around which the Higgs mass becomes tachyonic. 
We define the critical temperature $T_c$ around which the thermal mass becomes comparable to the zero temperature mass $g^2 M^2$ and $\phi_{\rm min}$ approaches $\phi_c$:
\begin{align}
T_c\equiv \frac{M}{\sqrt{\alpha}}. 
\end{align}
For the efficient transition of the instanton effect to make the periodic potential at this time, the temperature should be smaller than the condensation scale
$T_c\lesssim\Lambda_c$,
\footnote{If the relaxion is in the thermal equilibrium with a heat bath at a very high 
temperature $T$, the relaxion is fluctuated with $\delta\phi\sim T$. To avoid the transition over the periodic potential by this fluctuation, we require $\alpha g^2T^2\delta\phi^2\sim\alpha g^2T^4\lesssim\Lambda_c^3\braket{h}$ at $T\simeq T_c$, which reduces to 
$\frac{g^2}{\alpha}M^4\lesssim \Lambda_c^3\braket{h}$.
However, this condition is weaker than $T_c \lesssim \Lambda_c$.
}
which gives a lower limit for the dimensionless coupling as
\beq\label{eq:cond_al}
\alpha\gtrsim\lrp{\frac{M}{\Lambda_c}}^2 \gtrsim 1.
\eeq
Otherwise, the relaxion does not settle down to the local minimum where the electroweak scale is dynamically realized through the periodic potential.
For the success of the Big Bang Nucleosynthesis (BBN), the critical temperature needs to be larger than ${\rm MeV}$, giving a upper bound on the dimensionless parameter as
\beq\label{eq:al_upp}
\alpha\lesssim\lrp{\frac{M}{{\rm MeV}}}^2.
\eeq

To fix the field value of the relaxion by the periodic term, 
the slope of the periodic potential should be comparable to the linear term:
\beq\label{Acond}
M^3g\simeq \frac{\Lambda_c^3\braket{h}}{f}\left|\sin\lrp{\frac{\phi_c}{f}}\right|,
\eeq
where $\braket{h}$ is the electroweak VEV of the Higgs field.

Since the relaxion dynamically moves following the minimum of the potential (\ref{Tdom}) as temperature drops, the relaxion has kinetic energy with $(\dot{\phi}_{\rm min})^2\sim\lrp{M^3/(g\alpha M_{\rm pl})}^2$. For the relaxion to be stopped by the periodic potential, the kinetic energy should be smaller than $\Lambda_c^3\braket{h}$ as
\beq\label{eq:cond_veloci}
\frac{1}{g^2\alpha^2}\frac{M^6}{M_{\rm p}^2}\lesssim\Lambda_c^3\braket{h}.
\eeq

We have obtained four conditions on the relaxion parameters as (\ref{eq:cond_al}),~(\ref{eq:al_upp}),~(\ref{Acond}), 
and (\ref{eq:cond_veloci}), which are summarized as
\beq\label{eq:sum_cond}
\left\{
\begin{aligned}
&\lrp{\frac{M}{\Lambda_c}}^2\lesssim\alpha\lesssim\lrp{\frac{M}{\rm MeV}}^2,\\
&\alpha\gtrsim\frac{1}{g}\frac{M^3}{M_{\rm pl}\lrp{\Lambda_c^3\braket{h}}^{1/2}},\\
&g\simeq\frac{\Lambda_c^3\braket{h}}{M^3f}
\left|\sin\lrp{\frac{M}{gf}}\right|.
\end{aligned}
\right.
\eeq
Note that we can rewrite the last equation as
$1 \sim \left(\frac{\Lambda_c}{M}\right)^4 
\left(\frac{\langle h\rangle}{\Lambda_c}\right)
\left(\frac{M}{gf}\right) \left|\sin\lrp{\frac{M}{gf}}\right|$.
Since $M \gtrsim \Lambda_c$ and $\Lambda_c \sim \langle h \rangle$, 
the argument of the sine function in the relation needs to be larger than unity to have a solution. By an approximation $|\sin \left(\frac{M}{gf}\right)|\simeq1$, we can rewrite the last equation as
\beq\label{eq:appro_rela}
g\simeq \lrp{\frac{\Lambda_c}{M}}^{3} \lrp{\frac{\braket{h}}{f}}.
\eeq
The condition in the first line of (\ref{eq:sum_cond}) is independent of any choice of $g$ and $f$, and it requires $\alpha$ to be larger than unity. A large value of $\alpha$ could be achieved in the case with the large numbers of the species mediating between the relaxion and the thermal bath, or
$g^2 \alpha$ would be nothing to do with $g$ (or $M$) essentially.
With any value of $M$, we should require at least
\beq\label{eq:lowest_al}
\alpha\simeq\lrp{\frac{M}{\Lambda_c}}^2.
\eeq
This lowest hierarchy is achieved for the decay constant 
$f/M_{\rm pl}\lesssim\lrp{\braket{h}/\Lambda_c}^{3/2}\lrp{\Lambda_c/M}^{4}$,
which is obtained by substitution of  (\ref{eq:appro_rela}) and (\ref{eq:lowest_al}) into the condition in the second line of (\ref{eq:sum_cond}).

Using the approximated relation (\ref{eq:appro_rela}), we can rewrite the condition in the third line of (\ref{eq:sum_cond}) (corresponding to (\ref{eq:cond_veloci})) as
\beq\label{eq:cond_veloci_v2}
\alpha\gtrsim\lrp{\frac{M}{\Lambda_c}}^6\frac{f}{M_{\rm pl}}\lrp{\frac{\Lambda_c}{\braket{h}}}^{3/2}.
\eeq
Since $\alpha$ is limited from above as shown in the first line of (\ref{eq:sum_cond}), (\ref{eq:cond_veloci_v2}) gives a upper limit on the decay constant as
\beq\label{eq:cond_f}
\frac{f}{M_{\rm pl}}\lesssim \lrp{\frac{\braket{h}}{\Lambda_c}}^{3/2}\lrp{\frac{\Lambda}{\rm MeV}}^2
\lrp{\frac{\Lambda_c}{M}}^{4}.
\eeq
%

\subsection{The case (B): $T_{\rm R}<\Lambda_c$}
In the case where $T_{\rm R}<\Lambda_c$, 
we consider that the energy of inflaton oscillation 
dominates the Universe even when the relaxion reaches the critical point $\phi_c$ 
at which the Higgs mass vanishes. In this case, the temporal minimum of the relaxion is determined by the Hubble induced mass term as~(\ref{Hdom}) instead of the thermal one. When $H=H_c\sim (g/\sqrt{c})M$, the temporal minimum reaches $\phi=\phi_c$. At this point, the minimization condition $\pd_\phi V=0$ reads
\begin{align}\label{eq:relation_H}
gf\simeq \left( \frac{\Lambda_c}{M}\right)^3\braket{h}\sin\lrp{\frac{\phi_c}{f}}.
\end{align} 
The quarks of the SM need to obtain their observed masses for the successful BBN. 
The temperature at $H=H_c$ should be larger than $\mathcal{O}(1){\rm MeV}$ around which the BBN starts. This requirement gives an upper bound on the dimensionless parameter $c$  as
\beq
c<
10^{14}\times\lrp{\frac{g_{\ast}}{100}}^{-1}\lrp{\frac{g}{10^{-20}}}^2\lrp{\frac{M}{10{\rm TeV}}}^2.
\eeq

As discussed in the case of (A), the dynamical change of the potential minimum
gives kinetic energy for relaxion. In present case, the deformation of the potential is due to the decrease of the Hubble expansion rate. To stop the relaxion by the periodic potential, this kinetic energy should be smaller than the scale of the  potential $\Lambda_c^3\braket{h}$. Now since the potential minimum of the relaxion is determined by the Hubble induced mass as $\phi_{\rm min}\sim gM^3/(cH^2)$, the kinetic energy when the relaxion mechanism takes place at $H\sim H_c$ is estimated as $(\dot{\phi}_{\rm min})^2\sim M^4/c.$ Thus, we obtain a condition on the dimensionless parameter $c$ as
\beq\label{eq:cond_c}
c> \left(\frac{M}{\Lambda_c}\right)^4\left(\frac{\Lambda_c}{\braket{h}}\right) \gtrsim 1.
\eeq
Thus, to achieve the relaxion mechanism with the low reheating temperature, the coefficient of the Hubble induced mass term needs to be much larger than unity. 
As discussed in Sec.~\ref{model}, one of the possibilities to realize it is that the shift symmetry is broken by quantum gravity effects.
If this is the case, $c\gg 1$ implies that the cutoff scale of such effects would 
be smaller than $M_{\rm pl}$.
\\

In this section we have discussed the dynamics of the relaxion especially focusing on the time when the relaxion reaches the critical point $\phi_c = M/g$,
and then obtained the constraints of (\ref{eq:sum_cond}) 
for $T_R > \Lambda_c$,
and them of (\ref{eq:relation_H}) and (\ref{eq:cond_c}) for $T_R < \Lambda_c$.

In the case where 
$T_R>\Lambda_c$, the thermal mass term stabilizes the relaxion field. 
Then, a large $\alpha$ is required.
For example, the scale $M$ is around $10~{\rm TeV}$ and $\Lambda_c\sim 100$ GeV~(\ref{eq:scale}), the hierarchy is required with $\alpha_{\rm min}\sim10^4$. In this case, the relaxion mechanism is realized e.g by $(g,\,f)\sim(10^{-13},\,10^{9}\,{\rm GeV})$. With larger hierarchy, the relaxation
is realized with larger decay constants. For examples, $( g,~f)\sim(10^{-16},10^{12}\,{\rm GeV}),\,(10^{-20},10^{16}\,{\rm GeV})$. In the case where 
$T_R<\Lambda_c$, the Hubble induced mass stabilizes the relaxion field.
In this case, a sizable $c$ is required.
For $M\sim10~{\rm TeV}$, the dimensionless parameter $c$ needs to be larger than $10^8$.

\section{Conclusion and discussion}\label{concl}

Recently the relaxion mechanism to solve the Higgs hierarchy problem was proposed~\cite{Graham:2015cka}. 
In this paper, we have studied 
whether the mechanism is accomplished with the large Hubble expansion rate of inflation or high reheating temperature. Unlike the original one, we have discussed the scenario that the relaxion mechanism takes place after inflation, and 
then the Higgs VEV settles down to the electroweak scale.

To achieve the scenario, we have taken into account the effects of the 
Hubble induced mass or thermal one on the dynamics of the relaxion. In (\ref{Hmass}) and (\ref{eq:Tmass}), we have defined the masses with dimensionless parameters $c$ and $\alpha$. Then, by discussing the cosmological scenario of the relaxion, we have obtained the constraints on the parameters as  
(\ref{eq:cond_al}) and (\ref{eq:cond_c}).
From these constraints we can see that the dimensionless parameters need to be larger than unity for $M>\Lambda_c$. Therefore, to accomplish the relaxion mechanism with the Hubble induced mass or thermal one, 
there should exist additional shift symmetry breaking terms
in different ways from those given by the combination of $g \phi/M$.

To complete our scenario, it is necessary to consider UV completions 
and also the phenomenological aspects of the model. 
There exist several origins of the shift symmetry breaking
relevant to a sizable Hubble induced mass and thermal one.
It is also important to realize a large field excursion of the relaxion
\cite{Choi:2015fiu,Kaplan:2015fuy,Ibanez:2015fcv}. The relaxion can be a part of the dark matter component. Along the line of Ref.~\cite{Kobayashi:2016bue}, 
it is necessary to discuss whether the relic abundance of relaxion in our scenario can be consistent with the present data. 
The testability in collider physics would also be interesting as, for instance, in Refs.~\cite{Graham:2015cka,Batell:2015fma}.

Finally, we comment on the breaking of the (continuous)
shift symmetry of the relaxion before or during inflation.
Throughout this paper, we have assumed that 
an approximate continuous global symmetry is  spontaneously broken down
before or during inflation and the relaxion exists then as its Nambu-Goldstone mode.
(Note that isocurvature perturbation during inflation is suppressed 
in the presence of a sizable
Hubble induced mass in our scenario \cite{Dine:2004cq}.)
In contrast, if the spontaneous symmetry breaking occurs after inflation, 
the relaxion scenario might lead to contradictions with the observed Universe.
After the symmetry breaking,
the relaxion field would take random values in each patch of the Universe. 
The random values can be much larger than the scale $M/g$ owing to 
an almost flat potential with monodromy $g\phi/M$. Since we have discussed the relaxion dynamics based on an effective field theory, we can not control the higher terms of the relaxion potential (\ref{lin}) for $\phi \gg M/g$, where there might exist local minima,
and the relaxion would be trapped in them.
For discussion of this issue, we have to determine a UV theory. Further, 
by the symmetry breaking, 
topological defects would be formed and might cause serious problems for cosmology such as the domination of the domain walls even if there is a bias.


\section*{Acknowledgement}
This work is supported by MEXT-Supported Program for the Strategic Research Foundation at Private Universities,``Topological Science, Grant Number S1511006 
(T.H, N.T. and Y.Y.) and JSPS KAKENHI Grant Number 26247042 (T.H.).

  \end{document}